\documentclass[a4paper]{emulateapj}
\usepackage{hyperref}

\slugcomment{Accepted for publication in ApJ Letters}
\shorttitle{Stochastic orbital decay of MBH pairs}
\shortauthors{Fiacconi et al.}

\begin{document}

\title{Massive black hole pairs in clumpy, self-gravitating circumnuclear disks:\\stochastic orbital decay}
\author{Davide Fiacconi\altaffilmark{1,3}, Lucio Mayer\altaffilmark{1,3}, Rok Ro\v{s}kar\altaffilmark{1}, and Monica Colpi\altaffilmark{2,3,4}}
\email{fiacconi@physik.uzh.ch}
\altaffiltext{1}{Institute for Theoretical Physics, University of Zurich, Winterthurerstrasse 190, CH-8057 Z\"{u}rich, Switzerland}
\altaffiltext{2}{Dipartimento di Fisica ``G. Occhialini'', Universit\`a di Milano-Bicocca, Piazza della Scienza 3, I-20126 Milano, Italy}
\altaffiltext{3}{Kavli Institute for Theoretical Physics, University of California, Santa Barbara, CA 93106-4030, USA}
\altaffiltext{4}{Istituto Nazionale di Fisica Nucleare, Sezione di Milano-Bicocca, Piazza della Scienza 3, I-20126 Milano, Italy}


\begin{abstract}
We study the dynamics of massive black hole pairs in clumpy gaseous circumnuclear disks.
We track the orbital decay of the light, secondary black hole $M_{\bullet2}$ orbiting around the more massive primary at the center of the disk, using $N$-body/smoothed particle hydrodynamic simulations. 
We find that the gravitational interaction of $M_{\bullet2}$ with massive clumps $M_{\rm cl}$  erratically  perturbs the otherwise smooth orbital decay.
In close encounters with massive clumps, gravitational slingshots can kick the secondary black hole out of the disk plane.
The black hole moving on an inclined orbit then experiences the weaker dynamical friction of the stellar background, resulting in a longer orbital decay timescale. 
Interactions between clumps can also favor orbital decay when the black hole is captured by a massive clump which is segregating toward the center of the disk.  
The stochastic behavior of the black hole orbit emerges mainly when the ratio $M_{\bullet2}/M_{\rm cl}$ falls below unity, with decay timescales ranging from $\sim1$ to $\sim50$ Myr.
This suggests that describing the cold clumpy phase of the inter-stellar medium in self-consistent simulations of galaxy mergers, albeit so far neglected, is important to predict the black hole dynamics in galaxy merger remnants.
\end{abstract}



\section{Introduction}

Close massive black hole (MBH) pairs  with separations $\lesssim10$ pc are expected to form as a consequence of galaxy mergers during cosmic evolution (e.g. \citealt{begelman+80}; \citealt{mayer+07}; \citealt{chapon+13}).
When the galaxies involved in the merger are gas-rich, the pair is usually embedded in a gaseous circumnuclear disk (CND) guiding the BH inspiral down to (sub) parsec scales (\citealt{escala+05}; \citealt{dotti+07}, \citeyear{dotti+09}; \citealt{mayer+07}).

Massive gaseous CNDs form  at the center of merging remnants, following gas inflows driven by gravitational torques (\citealt{dimatteo+07}; \citealt{mayer+07}, \citeyear{mayer+10}; \citealt{beletsky+11}).
However, the same torques can also compress the gas and induce episodes of massive star formation (SF; i.e., merger-driven nuclear starbursts, \citealt{downes+98}; \citealt{smith+07}; \citealt{smirnova+10}).
In star forming regions the inter-stellar medium (ISM) is expected to be clumpy on scales of a few to several tens of parsecs, these being the typical sizes of molecular clouds (MCs, \citealt{klessen+10}).
Gravoturbulent fragmentation at scales of a few tens of parsecs, governed by the Toomre instability in a self-gravitating gaseous disk, is one of the processes that likely seeds MCs (e.g., \citealt{agertz+09a}; \citealt{tasker+09}).

Previous investigations probed the BH pair dynamics assuming that the self-gravitating gas in the CND is warm enough to avoid evolving into an unstable inhomogeneous phase.
However, a CND undergoing a starburst generally hosts a complex and inhomogeneous multi-phase ISM.
\citet{escala+05} briefly explored, in 3D simulations, the BH dynamics in inhomogeneous CNDs excited by an isothermal equation of state and noted that violent interactions with clumps can occur, but that in most cases the BHs end up at the center.
On smaller scales ($\lesssim0.1$ pc), \citet{lodato+09} investigated the effects of disk instability and SF onto the evolution of BH binaries using time-dependent 1D models, finding that orbital decay is very inefficient when SF depletes the disk background gas.
Albeit different in terms of both regime and technique, these two works suggest that it is not trivial to predict the outcome of the BH binary/pair shrinking process in a gaseous environment when the complexity of ISM is somehow taken into account.

The relevance of a clumpy ISM to galaxy formation as a whole has recently grown following the discovery that galaxies at $z\gtrsim2$ are clumpy at relatively large scales.
These clumps have typical mass $\gtrsim10^{8}$ $M_{\odot}$ and size of $\sim1$ kpc and are believed to be triggered by the Toomre instability in massive, rapidly accreting disks at high redshift (\citealt{genzel+06}; \citealt{agertz+09b}; \citealt{dekel+09}; \citealt{ceverino+12}).
While we do not model such high-$z$ massive clumps in this work, our results may give a hint on the decay process of MBH pairs in such larger-scale clumpy high-$z$ galactic disks.

In this Letter, we compare the evolutions of MBH pairs in smooth and highly inhomogeneous CNDs by means of $N$-body/smoothed particle hydrodynamic (SPH) simulations.
We explore the effect of the lumpiness of the ISM in an idealized and controlled way by varying the degree of energy dissipation in the gas.
We choose this approach because modeling the multi-phase ISM properly is computationally demanding as it requires including gas cooling both in the optically thin and thick regime, SF and feedback.
Relatively low BH mass ratios, between 0.05 and 0.2, are explored in order to magnify the effect of the disk lumpiness on the dynamics of the light, secondary BH.
We show that the behavior can be highly stochastic, often delaying but in some cases also favoring the pairing of the BHs.
In Section \ref{sec2} we describe the numerical simulations, and in Section \ref{sec3} we present the results.
Section \ref{sec4} contains the discussion and our conclusions.


\section{Numerical simulations}\label{sec2}

We perform a suite of numerical simulations to study the evolution of a MBH pair embedded in a gaseous CND which, in turn, is at the center of a stellar spheroid.
All the simulations were run with the TreeSPH $N$-body code {\tt GADGET2} \citep{springel+05}.
Our models resemble the inner region of the remnant of a merger that involved two galaxies with a central MBH each, in a fashion similar to \citet{escala+05} and \citeauthor{dotti+07} (\citeyear{dotti+07}, \citeyear{dotti+09}).
The gaseous disk has a Mestel surface density profile with a scale length $R_{\rm d}=100$ pc and a radial extent of $\sim150$ pc.
The vertical structure is initially Gaussian, with a scale height $z_{\rm d}(R)=h\,R$ and aspect ratio $h=0.05$.
The gas has an initial uniform temperature $T_0=20,000$ K.
The disk is embedded in a Plummer stellar spheroid that represents the innermost part of a bulge, with scale radius $r_{\star}=50$ pc and radial extent of $\sim500$ pc.
Among the different runs, we vary the mass of the spheroid $M_{\star}$ and the mass of the disk $M_{\rm d}$, but we fix the ratio $M_{\star}/M_{\rm d}=5$, in fair agreement with observations \citep{downes+98}.
We place the first BH of mass $M_{\bullet1}=10^{7}$ $M_{\odot}$ at the center of the disk and we let the models relax for 10 Myr toward equilibrium.
We assume a polytropic equation of state for the gas with adiabatic index $\gamma=1.4$, in agreement with previous theoretical (\citealt{klessen+07}; \citealt{mayer+07}, \citeyear{mayer+10}) and observational \citep{downes+98} works that studied the conditions of the central gas in ongoing mergers or merger remnants.

Then, we finally added the second BH of mass $M_{\bullet2}=q\,M_{\bullet1}$ at the initial separation $a_0=60$ pc from the central one.
Both BHs are treated as collisionless particles.
The orbit of the secondary BH is specified by the ratio $f$ between the radial and the azimuthal components of the initial velocity $\mathbf{v}_{0}$, with the constraint $|\mathbf{v}_{0}|=V_{\rm c}(a_{0})$, where $V_{\rm c}(a_{0})$ is the circular velocity in  $a_{0}$.
$f$ also specifies the initial eccentricity $e_0$ of the orbit, $e_0\sim\sqrt{1-1/(1+f^2)}$.
All the models are composed of $2\times10^5$ SPH particles and $1\times10^{6}$ collisionless particles for the gaseous disk and the stellar spheroid, respectively.
This corresponds to a mass resolution $m_{\rm p}$ that varies between 500 and 2500 $M_{\odot}$, depending on the mass of the disk and the spheroid.
The force resolution set by the gravitational softening is $\epsilon_{\rm g}=0.5$ pc for all the particles.

For each choice of parameters $M_{\rm d}$, $q$ and $f$, we initialize three sets of initial conditions.
In the first one, which we will refer to as the ``smooth'' set, the secondary BH is added immediately after the 10 Myr relaxation phase.
In the other two, we added the secondary BH after we forcefully make the disk Toomre unstable by cooling it during an additional relaxation phase via the phenomenological cooling term:
\begin{equation} \label{eq:cooling}
\Lambda_{\rm cool} =-\frac{u}{t_{\rm cool}},
\end{equation}
where $u$ is the specific internal energy of the gas and $t_{\rm cool}$ is a constant cooling timescale.
We chose different parameters to create two different environments: in the first case (``clumpy 1''), we further relaxed the disk for 2 Myr with $t_{\rm cool}=0.2$ Myr.
The disk becomes violently unstable and fragments in many small and dense clumps that account for $\gtrsim M_{\rm d}/2$.
In the second case (``clumpy 2'') we used $t_{\rm cool}=1$ Myr for 10 additional Myr and we turn off the cooling locally for densities $>5\times10^{5}$ H cm$^{-3}$.
The disk has a more developed background component compared to ``clumpy 1'' models, with fewer, less concentrated clumps.

Table \ref{table1} lists all the simulations and summarizes the adopted parameters.
The values of $t_{\rm cool}$ for both ``clumpy 1'' and ``clumpy 2'' models are chosen to be comparable to the dynamical time of the disks at $\sim R_{\rm d}$ to allow gravoturbulent fragmentation to set in and sustain.
We keep the density switch for the cooling during ``clumpy 2'' simulations even after we inserted the secondary BH.
We stress that Equation (\ref{eq:cooling}) represents a phenomenological way to create and maintain strong inhomogeneities in the CND.
\begin{deluxetable}{l c c c c c}
\tablecaption{List of performed simulations and of their parameters.\label{table1}}
\tablehead{
\colhead{Label} & \colhead{$M_{\rm d}$} & \colhead{$q$\tablenotemark{a}} & \colhead{$f$} & \colhead{$e_{0}$\tablenotemark{b}} & \colhead{$t_{\rm cool}$} \\
\colhead{} & \colhead{$(M_{\odot})$} & \colhead{} & \colhead{} & \colhead{} & \colhead{(Myr)}
}
\startdata
q005f02LM & $10^{8}$          & 0.05 & 0.2  & 0.2  & 1.0 \\
q005f1LM  & $10^{8}$          & 0.05 & 1.0  & 0.7  & 1.0 \\
q02f025LM & $10^{8}$          & 0.2  & 0.25 & 0.25 & 1.0 \\
q02f2LM   & $10^{8}$          & 0.2  & 2.0  & 0.9  & 1.0 \\
q01f02HM  & $5 \times 10^{8}$ & 0.1  & 0.2  & 0.2  & 0.5 \\
q01f2HM   & $5 \times 10^{8}$ & 0.1  & 2.0  & 0.9  & 0.5 \\
q02f02HM  & $5 \times 10^{8}$ & 0.2  & 0.2  & 0.2  & 0.5 \\
q02f2HM   & $5 \times 10^{8}$ & 0.2  & 2.0  & 0.9  & 0.5 
\enddata
\tablenotetext{a}{$q=M_{\bullet2}/M_{\bullet1}$, $M_{\bullet1}=10^7$ $M_{\odot}$.}
\tablenotetext{b}{$e_0\sim\sqrt{1-1/(1+f^2)}$.}
\end{deluxetable}


\section{Results}\label{sec3}


\subsection{Orbital Decay in a Smooth Disk: Overview}\label{subsec3.1}

\begin{figure*}
\plotone{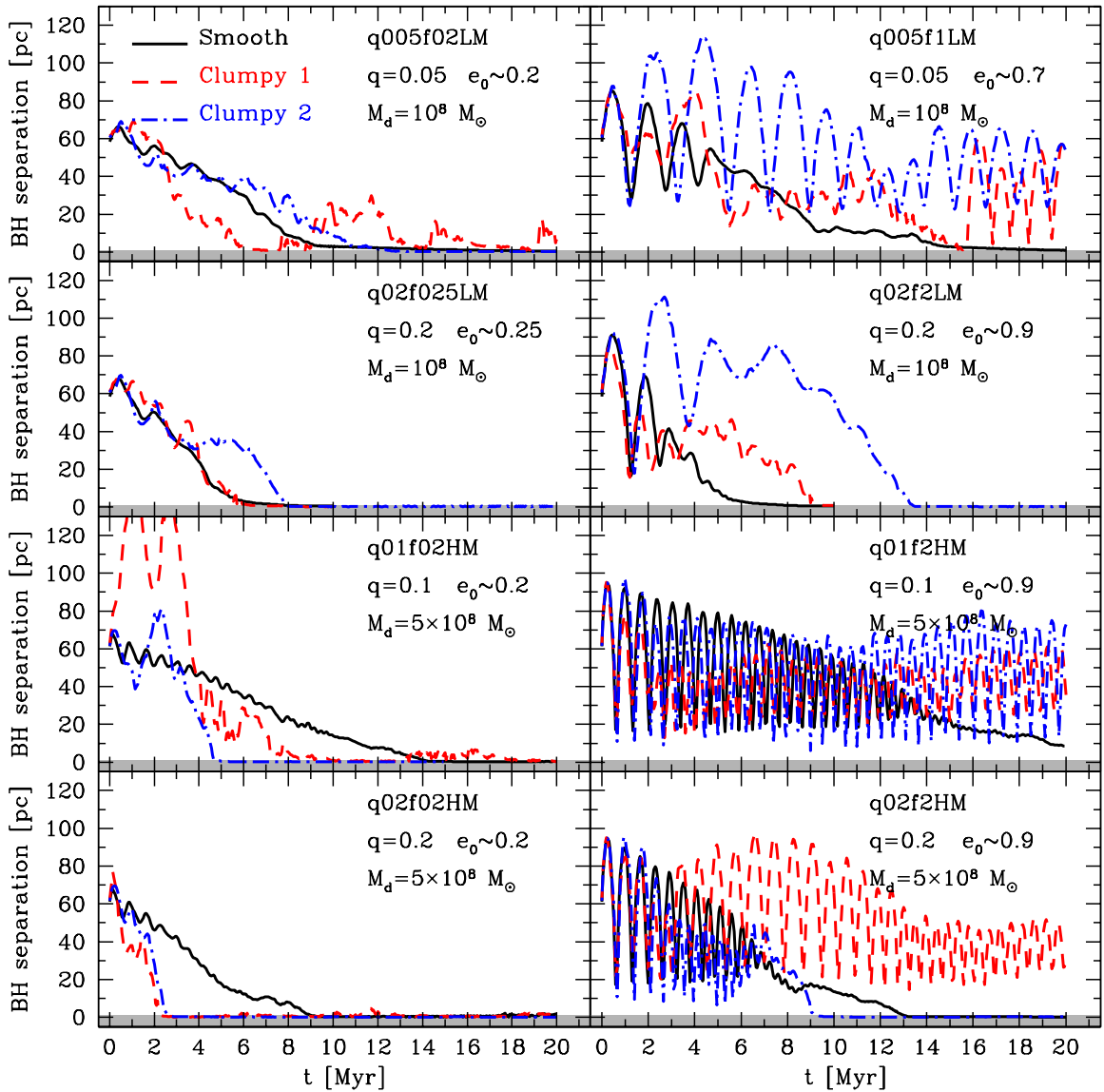}
\caption{
Time evolution of the BH separation for all the corresponding ``smooth'', ``clumpy 1'' and ``clumpy 2'' models.
The top two rows show runs with $M_{\rm d}=10^{8}$ $M_{\odot}$, whereas the bottom two with $M_{\rm d}=5\times10^{8}$ $M_{\odot}$.
For each pair of rows, the first and the second one shows runs with lower and higher $q$, respectively.
The left and the right column shows runs with lower and higher $e_{0}$, respectively. 
Black continuous lines, red dashed lines and blue dot-dashed lines show the BH separation of the ``smooth'', ``clumpy 1'' and ``clumpy 2'' models, respectively.\label{fig1}}
\end{figure*}

\begin{figure}
\plotone{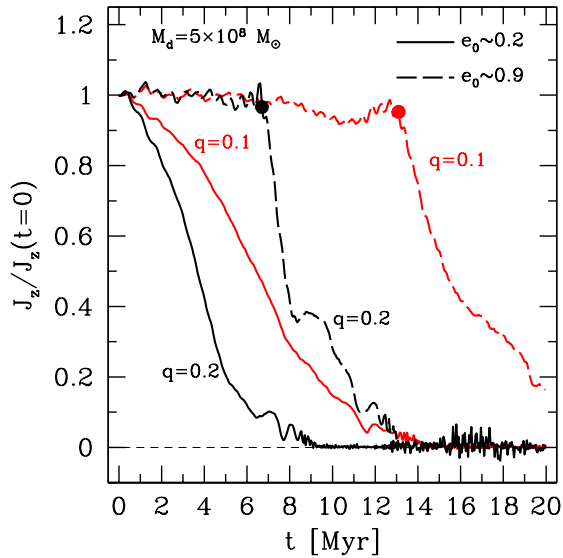}
\caption{Time evolution of the secondary BH angular momentum for the ``smooth'' models with $M_{\rm d}=5\times10^{8}$ $M_{\odot}$.
Solid and dashed lines represent runs with $e_{0}\simeq0.2$ and $e_{0}\simeq0.9$, respectively. 
Black and red colors represent runs with $q=0.1$ and $q=0.2$, respectively. 
The dots qualitatively mark the transition between the two phases described in the text.\label{fig2}}
\end{figure}

The secondary BH of all the ``smooth'' simulations moves toward the disk center on a typical timescale of $\sim10$ Myr.
This is shown in Figure \ref{fig1}, which compares the time evolution of the separation between the two BHs for all the corresponding ``smooth'', ``clumpy 1'' and ``clumpy 2'' simulations.
The orbital decay of the secondary BH in the ``smooth'' disks is characterized by two phases.
Initially, the BH induces a trailing hydrodynamical wake that, in turn, makes the orbit circularize \citep{dotti+07} because of conventional dynamical friction (DF; \citealt{chandra+43}; \citealt{ostriker+99}; \citealt{colpi+99}).
When the orbit is close to circular, the relative velocity between the BH and the gas becomes low.
This causes a change in the orbital decay timescale.
The BH induces a density wave perturbation, eventually amplified by the self-gravity of the disk that exerts a global torque on the perturber itself, initiating its rapid sinking toward the center.
The latter phase resembles the regime of standard Type I  planet migration \citep{lin+86}.
An aspect that has not been appreciated in the literature is that the secondary BH's angular momentum loss occurs on an intrinsically shorter timescale during the latter phase, as shown in Figure \ref{fig2}.

Furthermore, while DF shuts off at separations such that the enclosed mass within the pair's orbit is $\sim M_{\bullet1}+M_{\bullet2}$, global torques will continue to act as long as a sufficiently massive gaseous disk is present at large radii and the BH does not open a gap \citep{chapon+13}.
Finally, in addition to the disk contribution to the orbital decay, the secondary BH also suffers DF exerted by the stellar spheroid.
However, the stellar torque is weaker than the torque that comes from the gaseous disk, which mainly drives the dynamical evolution of the pair.


\subsection{BH Pair Evolution in a Clumpy Disk}

\begin{figure}
\plotone{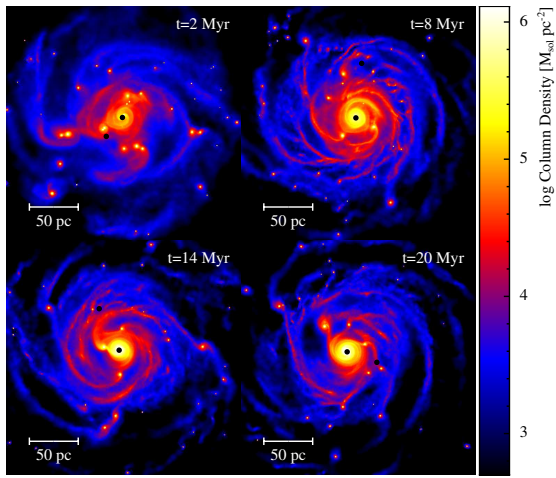}
\caption{Time evolution of the gas surface density of the ``clumpy 1'' model of run q02f2HM.
The size of each panel is 240 pc per edge.
The position of the BHs are marked by two black dots.\label{fig3}}
\end{figure}

\begin{figure*}
\plotone{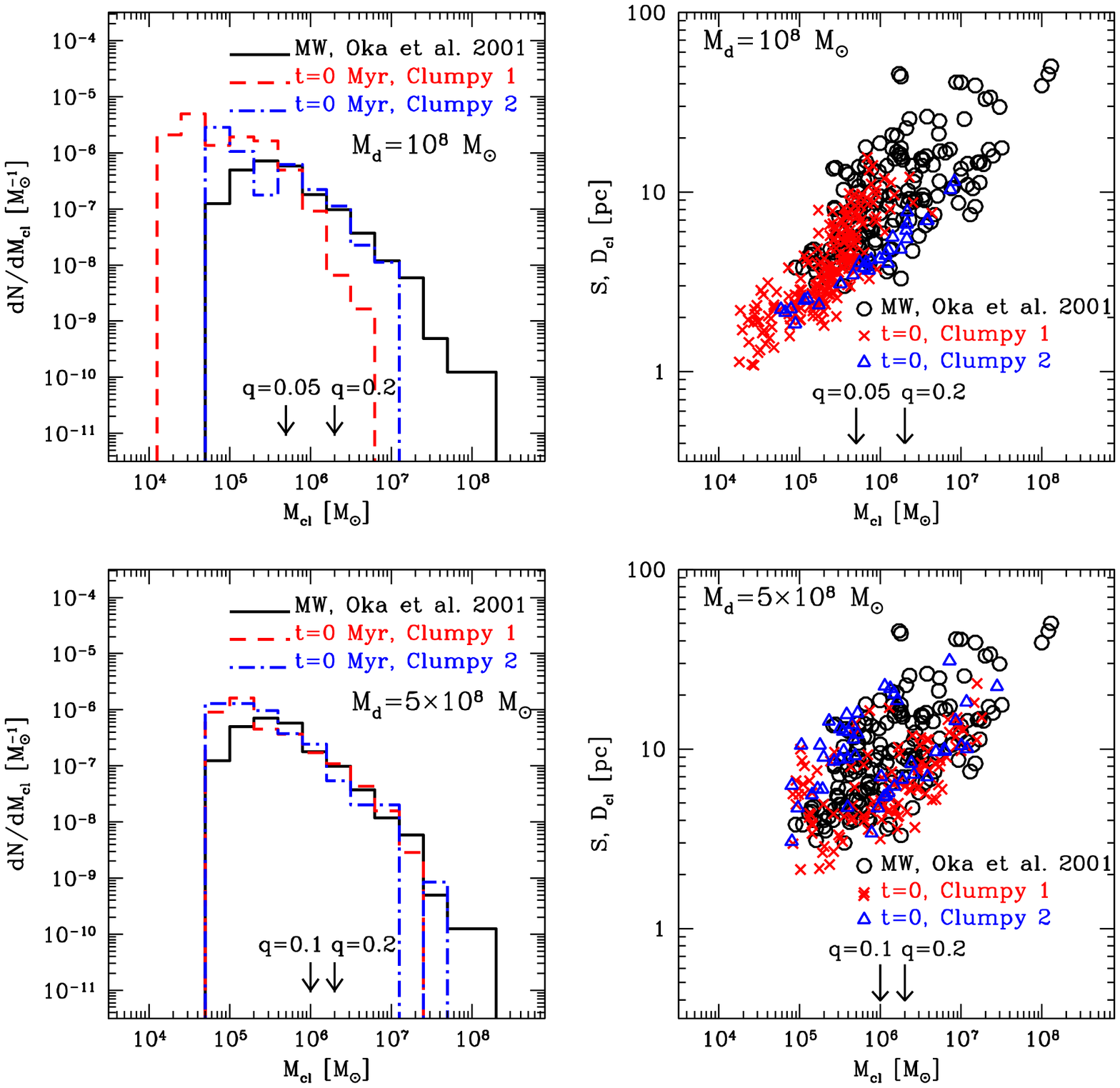}
\caption{Top-left panel: normalized mass function of clumps for ``clumpy 1'' and ``clumpy 2'' models with $M_{\rm d}=10^{8}$ $M_{\odot}$.
Bottom-left panel: same as before for models with $M_{\rm d}=5\times10^{8}$ $M_{\odot}$.
Top-right panel: clump mass vs. size ($S$ from \citet{oka+01} and clump diameter $D_{\rm cl}$) for models with $M_{\rm d}=10^{8}$ $M_{\odot}$.
Bottom-right panel: same as before for models with $M_{\rm d}=5\times10^{8}$ $M_{\odot}$.
Black continuous lines (circles), red dashed lines (crosses) and blue dot-dashed lines (triangles) show the data from \citet{oka+01}, ``clumpy 1'' and ``clumpy 2'' models, respectively.
Vertical arrows indicate the masses of the secondary BHs.\label{fig4}}
\end{figure*}

The orbital evolution of the secondary BH in a clumpy disk is affected by the dynamical interaction with clumps acting as massive perturbers.
Figure \ref{fig3} shows the gas surface density of the disk for the ``clumpy 1'' model of run q02f2HM.
The disk is sprinkled with massive clumps and non-axisymmetric features that exert torques on the secondary BH.
Figure \ref{fig4} shows the normalized initial mass function (just after the relaxation phase) and the mass-size relation of the clumps of ``clumpy 1'' and ``clumpy 2'' models.
We compare our data with the observed sample of MCs in the Galactic center taken from \citet{oka+01}.
Although the Galactic center likely represents a less extreme environment than the inner region of a merger remnant that we aim at modeling, our models are in fair agreement with the observational data and are only weakly influenced by the value of $t_{\rm cool}$ adopted in the relaxation phase.

Close encounters (at separation of $\sim$3-10 pc) between the secondary BH and the massive clumps occur in both ``clumpy 1'' and `` clumpy 2'' models.
BH-clump interactions act as gravitational slingshots, causing an exchange of orbital energy and angular momentum.
These impulsive perturbations make the BH deviate from its original orbit, shifting its mean orbital radius either inward or outward.

Gravitational slingshots can delay the sinking of the secondary BH when they cause its temporary ejection from the disk plane.
This happens both with the dense clumps of ``clumpy 1'' models and with the more diffuse ones of ``clumpy 2'' models, but preferentially for high-$e_{0}$ runs (e.g. run q005f1LM, q02f2LM and q01f2HM).
After the interaction, the secondary BH moves on an eccentric orbit, tilted up to $\sim15^{\circ}$-$40^{\circ}$ with respect to the original plane for both ``clumpy 1'' and ``clumpy 2'' models.
The BH then spends most of the orbital period moving outside the disk, where is mostly subjected to DF against the stellar background and the decay timescale becomes longer, in agreement with \citet{escala+05}.
The actual value of the decay timescale can depend substantially on the density profile of the stellar spheroid.
Sinking BHs can stall in constant density cores when they approach the core radius, as shown by \citet{read+06}.

Clumps have also an indirect retarding effect on the BH orbital decay, especially for eccentric orbits.
They produce spiral arms that exert torques on the secondary BH and perturb the background component, stifling the coherence of the BH hydrodynamical wake that would lead to the circularization of the orbit and the subsequent efficient decay phase described in Section \ref{subsec3.1}.

On the other hand, either the BH can be captured inside a massive clump, or a clump can be tidally heated and then disrupted and accreted by the orbiting BH during a close passage at low relative velocity.
In both cases, the BH ends up embedded in a gaseous cloud bound to it.
Although this cloud cannot be actually accreted due to the lack of an accretion prescription in our simulations, it provides an higher effective mass that reduces the orbital decay timescale\footnote{We recall that both Type I migration and DF torques that provide loss of angular momentum and/or orbital energy scale with the perturber mass squared.} up to $\sim2$ Myr (e.g., ``clumpy 2'' model of run q01f02HM and q02f02HM).


\section{Discussion and conclusions}\label{sec4}

We compared the orbital decay of a MBH pair in clumpy and inhomogeneous gaseous CNDs with that in smooth disks by means of idealized $N$-body/SPH simulations.
We found that gravitational interactions between the BH and clumps do play a role in the dynamics of these pairs.
Gravitational slingshots frequently perturb the orbit of the moving BH, stochastically increasing or decreasing the separation of the pair.
Scattering between massive clumps and the secondary BH can eject the latter from the disk plane, having a net retarding effect on its orbital decay since in this case the dominant source of drag on the BH is the low density stellar background.
We can estimate this orbital decay timescale $\tau_{\rm decay}$ for a BH of mass $M_{\bullet}$ in circular motion inside a Plummer bulge of total mass $M_{\rm b}$ and scale radius $b$ using standard DF \citep{chandra+43} as $\tau_{\rm decay}\simeq\tau_{\rm decay}^{\rm(ext)}+\tau_{\rm decay}^{\rm(int)}$, where:
\begin{equation}
\tau_{\rm decay}^{\rm(ext)}\simeq\frac{0.5305}{\ln\Lambda}\,\left(\frac{M_{\rm b}}{M_{\bullet}}\right)\,\left[(r_{\rm in}/b)^{7/2}-1\right]\,t_{\rm dyn},
\end{equation}
is the time to move from an initial radius $r_{\rm in}$ to $\sim b$, and:
\begin{equation}
\tau_{\rm decay}^{\rm(int)}\simeq\frac{0.2172}{\ln\Lambda}\,\left(\frac{M_{\rm b}}{M_{\bullet}}\right)\,t_{\rm dyn},
\end{equation}
is the decay time inside the core.
As usual, $\ln\Lambda$ is the Coulomb logarithm and $t_{\rm dyn}=1/\sqrt{G\rho_0}$ is the dynamical time associated with central density $\rho_0$ of the Plummer profile.
We estimate that $\tau_{\rm decay}$ would vary typically between $\sim20$ and $\sim50$ Myr depending on the secondary-BH mass and the properties of the stellar spheroid that we assumed for our simulations.
We used $\ln\Lambda\sim5$ and $r_{\rm in}=60$ pc.
We also checked that the decay timescale would vary in the range $\sim$20-80 Myr if we adopted a (de-projected) S\'ersic profile for the bulge with structural parameters (mass, effective radius, S\'ersic index) ranging from those typical of massive classical bulges (e.g., Andromeda galaxy, \citealt{widrow+03}; \citealt{courteau+11}) to those of pseudobulges in late-type spirals (\citealt{fisher+08}; \citealt{fisher+09}).

At the same time, the BH pairing can be accelerated either when the secondary is scattered toward the disk center or it manages to accrete a bound gaseous cloud around it.
The second case occurs when the orbiting BH is captured inside a massive clump or when it rips mass off of a clump.
Then, the higher total mass of the BH+cloud system, compared to $M_{\bullet}$, can reduce the orbital decay timescale by up to $\sim2$ Myr.
However, BH accretion/feedback may influence the local distribution of mass around the secondary and, in turn, affect its sinking toward the center.

We estimate a threshold mass $\mathcal{M}_{\bullet}$ below which a $M_{\bullet}$ BH orbiting in a disk with mass $M_{\rm g}$ and scale radius $R$ will likely be scattered by clumps with $M_{\rm cl}>M_{\bullet}$.
The BH and the clumps both migrate toward the center on a timescale $\tau_{\bullet}$ and $\tau_{\rm cl}$, respectively.
We can envision that the BH moves in an almost steady-state environment in which massive clumps that dissolves close to the center are continuously replaced by new clumps on the same timescale, so that the number and the spatial distribution of massive clumps remains almost constant in time, on average.
The BH moves radially with an effective relative velocity $v_{\bullet}\sim R/\tau_{\bullet}\sim10$ km s$^{-1}$ with respect to the clumps.
This is a reasonable assumption when we focus on massive clumps only since they form because of the merging of smaller clumps during their migration on the timescale $\tau_{\rm cl}$ and dissolve when they reach the center on the same timescale.
Under these crude assumptions, we can imagine that the BH moves radially by a random walk and the condition for the scattering by clumps to be sizable reads:
\begin{equation} \label{eq:condition}
1\ll N\sim R^{2} n_{\rm cl}^{2}\sigma_{\rm cl}^{2}, 
\end{equation}
where $n_{\rm cl}\sim N_{\rm cl}/V_{\rm disk}$ is the number density of clumps in the disk volume $V_{\rm disk}\sim h\,R^3$ with aspect ratio $h$, and $\sigma_{\rm cl}$ is the cross section for a gravitational interaction with a clump.
Massive clumps have a size $\sim$5-10 pc, whereas the influence radius for interaction with the BH $r_{\rm g}\sim G M_{\rm cl}/v_{\bullet}^{2}\gtrsim10$ pc for $M_{\rm cl}\gtrsim10^{6}$ $M_{\odot}$, therefore we can estimate $\sigma_{\rm cl}\sim\pi r_{\rm g}^{2}$.
If we assume that $\tau_{\bullet}$ scales as for DF, $\tau\sim(\Theta/ \ln\Lambda)\,(M_{\rm g}/M_{\bullet})\,t_{\rm dyn}$, where $\Theta\sim1$, we can rewrite Equation (\ref{eq:condition}) in terms of the BH mass $\mathcal{M}_{\bullet}$ under which scatterings are relevant:
\begin{equation}
M_{\bullet}\ll\mathcal{M}_{\bullet}\sim\frac{\left(N_{\rm cl}\,h^{-1}\right)^{1/4}}{\ln\Lambda}\,\left(\frac{M_{\rm cl}}{M_{\rm g}}\right)^{1/2} M_{\rm g}.
\end{equation}

It is interesting to apply this order-of-magnitude estimate to high-$z$ galaxies with few clumps with mass of order of the maximum mass allowed by Toomre instability, $M_{\rm cl}\sim\eta^{2}M_{\rm g}$, where $\eta$ is the gas fraction of the system \citep{escala+08}:
\begin{equation}
\mathcal{M}_{\bullet}\sim7\times10^{7}\,\bigg(\frac{N_{\rm cl}}{4}\bigg)^{1/4}\bigg(\frac{\eta}{0.4} \bigg)\bigg(\frac{M_{\rm g}}{10^{9}\;{\rm M_{\odot}}}\bigg)\,{\rm M_{\odot}},
\end{equation}
where we assumed $h=0.2$ and $\ln\Lambda=5$.
For massive systems with $M_{\rm g}\sim10^{10}$ $M_{\odot}$, as observed a $z\gtrsim2$ (\citealt{genzel+06}; \citeyear{genzel+10}), this would lead to a threshold mass even a factor of $\gtrsim10$ higher, suggesting that, at least at high-$z$,  clump-BH interactions can be important for massive BHs, up to the mass of some of the largest found today in early-type galaxies ($\sim10^9$ $M_{\odot}$).

Despite this is only an order-of-magnitude calculation, it is significant to note that the dynamics of virtually all the MBH binaries that space-born gravitational wave detectors such as eLISA would probe at $z\gtrsim3$  (e.g. \citealt{amaro+13}) could be in principle affected by the clumpy environment in the early stages of pairing.
Indeed, although our controlled experiments do not change the overall scenario according to which MBH pairs in gaseous environments tend to evolve toward tight binary, the even higher inhomogeneity of the ISM in gas-rich, high redshift ($z\gtrsim$2-3) galaxies may reasonably lead to stronger interactions and ejections of MBHs, slowing down the decay by a factor $\gtrsim10$ (R. Ro\v{s}kar et al., in preparation).
If ejections causing delayed decay are common even at low redshift, close MBH pairs at separations in the range 10-100 pc should not be rare.
A first example of such a system might be the Seyfert galaxy NGC 3393 \citep{fabbiano+11}. 

Clearly, apart from being highly idealized, our models miss important physical ingredients (i.e., SF/feedback, BH accretion/feedback) that likely play a role in shaping the structure of the ISM.
Forthcoming simulations incorporating such mechanisms will allow us to better assess the magnitude of the effects illustrated in this Letter.
However, our work represents a first step towards elucidating the role of the structure of the ISM in the orbital decay of MBH pairs and binaries, these being pivotal sources of gravitational waves that will be targeted by future gravitational wave experiments.


\acknowledgments
We thank the anonymous referee for comments that helped the improvement of the Letter.
The simulations were performed with the \emph{Schr\"{o}dinger} and the \emph{ZBOX4} cluster at the University of Zurich.
We thank M. Dotti, J. Guedes, G. Lake, F. Meru, C. Miller and H. Perets for useful discussions.
D.F. and R.R. are supported by the Swiss National Science Foundation under grant \#No. 200021\_140645.




\begin{thebibliography}{}

\bibitem[Agertz et al.(2009a)]{agertz+09a} Agertz, O., Lake, G., Teyssier, R., et al.\ 2009a, \mnras, 392, 294

\bibitem[Agertz et al.(2009b)]{agertz+09b} Agertz, O., Teyssier, R., \& Moore, B.\ 2009b, \mnras, 397, L64

\bibitem[Amaro-Seoane et al.(2013)]{amaro+13} Amaro-Seoane, P., Aoudia, S., Babak, S., et al.\ 2013, GW Notes, Vol.~6, p.~4-110, 6, 4

\bibitem[Begelman et al.(1980)]{begelman+80} Begelman, M.~C., Blandford, R.~D., \& Rees, M.~J.\ 1980, \nat, 287, 307

\bibitem[Beletsky et al.(2011)]{beletsky+11} Beletsky, Y., Gadotti, D.~A., Moiseev, A., Alves, J., \& Kniazev, A.\ 2011, \mnras, 418, L6

\bibitem[Ceverino et al.(2012)]{ceverino+12} Ceverino, D., Dekel, A., Mandelker, N., et al.\ 2012, \mnras, 420, 3490

\bibitem[Chandrasekhar(1943)]{chandra+43} Chandrasekhar, S.\ 1943, \apj, 97, 255

\bibitem[Chapon et al.(2013)]{chapon+13} Chapon, D., Mayer, L., \& Teyssier, R.\ 2013, \mnras, 429, 3114

\bibitem[Colpi et al.(1999)]{colpi+99} Colpi, M., Mayer, L., \& Governato, F.\ 1999, \apj, 525, 720

\bibitem[Courteau et al.(2011)]{courteau+11} Courteau, S., Widrow, L.~M., McDonald, M., et al.\ 2011, \apj, 739, 20

\bibitem[Dekel et al.(2009)]{dekel+09} Dekel, A., Sari, R., \& Ceverino, D.\ 2009, \apj, 703, 785

\bibitem[Di Matteo et al.(2007)]{dimatteo+07} Di Matteo, P., Combes, F., Melchior, A.-L., \& Semelin, B.\ 2007, \aap, 468, 61

\bibitem[Dotti et al.(2007)]{dotti+07} Dotti, M., Colpi, M., Haardt, F., \& Mayer, L.\ 2007, \mnras, 379, 956

\bibitem[Dotti et al.(2009)]{dotti+09} Dotti, M., Ruszkowski, M., Paredi, L., et al.\ 2009, \mnras, 396, 1640 

\bibitem[Downes \& Solomon(1998)]{downes+98} Downes, D., \& Solomon, P.~M.\ 1998, \apj, 507, 615 

\bibitem[Escala et al.(2005)]{escala+05} Escala, A., Larson, R.~B., Coppi, P.~S., \& Mardones, D.\ 2005, \apj, 630, 152

\bibitem[Escala \& Larson(2008)]{escala+08} Escala, A., \& Larson, R.~B.\ 2008, \apjl, 685, L31

\bibitem[Fabbiano et al.(2011)]{fabbiano+11} Fabbiano, G., Wang, J., Elvis, M., \& Risaliti, G.\ 2011, \nat, 477, 431

\bibitem[Fisher \& Drory(2008)]{fisher+08} Fisher, D.~B., \& Drory, N.\ 2008, \aj, 136, 773

\bibitem[Fisher et al.(2009)]{fisher+09} Fisher, D.~B., Drory, N., \& Fabricius, M.~H.\ 2009, \apj, 697, 630

\bibitem[Genzel et al.(2006)]{genzel+06} Genzel, R., Tacconi, L.~J., Eisenhauer, F., et al.\ 2006, \nat, 442, 786

\bibitem[Genzel et al.(2010)]{genzel+10} Genzel, R., Tacconi, L.~J., Gracia-Carpio, J., et al.\ 2010, \mnras, 407, 2091

\bibitem[Klessen et al.(2007)]{klessen+07} Klessen, R.~S., Spaans, M., \& Jappsen, A.-K.\ 2007, \mnras, 374, L29 

\bibitem[Klessen et al.(2010)]{klessen+10} Klessen, R.~S., Glover, S.~C.~O., Clark, P.~C., et al.\ 2010, in AIP Conf. Proc, 1294, The First Stars and Galaxies: Challenges for the Next Decade (Melville, NY: AIP), Ed. Whalen, D.~J., Bromm, V., \& Yoshida, N., 28

\bibitem[Lin \& Papaloizou(1986)]{lin+86} Lin, D.~N.~C., \& Papaloizou, J.\ 1986, \apj, 309, 846

\bibitem[Lodato et al.(2009)]{lodato+09} Lodato, G., Nayakshin, S., King, A.~R., \& Pringle, J.~E.\ 2009, \mnras, 398, 1392

\bibitem[Mayer et al.(2007)]{mayer+07} Mayer, L., Kazantzidis, S., Madau, P., Colpi, M., Quinn, T., \& Wadsley, J.\ 2007, Science, 316, 1874

\bibitem[Mayer et al.(2010)]{mayer+10} Mayer, L., Kazantzidis, S., Escala, A., \& Callegari, S.\ 2010, \nat, 466, 1082 

\bibitem[Oka et al.(2001)]{oka+01} Oka, T., Hasegawa, T., Sato, F., et al.\ 2001, \apj, 562, 348

\bibitem[Ostriker(1999)]{ostriker+99} Ostriker, E.~C.\ 1999, \apj, 513, 252 

\bibitem[Read et al.(2006)]{read+06} Read, J.~I., Goerdt, T., Moore, B., et al.\ 2006, \mnras, 373, 1451

\bibitem[Smirnova \& Moiseev(2010)]{smirnova+10} Smirnova, A., \& Moiseev, A.\ 2010, \mnras, 401, 307

\bibitem[Smith et al.(2007)]{smith+07} Smith, B.~J., Struck, C., Hancock, et al.\ 2007, \aj, 133, 791

\bibitem[Springel(2005)]{springel+05} Springel, V.\ 2005, \mnras, 364, 1105

\bibitem[Tasker \& Tan(2009)]{tasker+09} Tasker, E.~J., \& Tan, J.~C.\ 2009, \apj, 700, 358 

\bibitem[Widrow et al.(2003)]{widrow+03} Widrow, L.~M., Perrett, K.~M., \& Suyu, S.~H.\ 2003, \apj, 588, 311

\end{thebibliography}
\end{document}